# Operation of a 1-Liter-Volume Gaseous Argon Scintillation Counter

K. Kazkaz,[1] M. Foxe,[2,1] A. Bernstein,[1] C. Hagmann,[1]
I. Jovanovic,[2] P. Sorensen,[1] W. S. Stoeffl,[1] and C. D. Winant[3,1]

[1]*Lawrence Livermore National Laboratory, Livermore CA 94550*
[2]*Purdue University, West Lafayette, IN 47907*
[3]*University of California, San Francisco, CA 94143*

(Dated: 2 November 2018)

## Abstract

We have built a gas-phase argon ionization detector to measure small nuclear recoil energies ($<$ 10 keVee). In this paper, we describe the detector response to X-ray and gamma calibration sources, including analysis of pulse shapes, software triggers, optimization of gas content, and energy- and position-dependence of the signal. We compare our experimental results against simulation using a 5.9-keV X-ray source, as well as higher-energy gamma sources up to 1332 keV. We conclude with a description of the detector, DAQ, and software settings optimized for a measurement of the low-energy nuclear quenching factor in gaseous argon. This work was performed under the auspices of the U.S. Department of Energy by Lawrence Livermore National Laboratory in part under Contract W-7405-Eng-48 and in part under Contract DE-AC52-07NA27344. Funded by Lab-wide LDRD. LLNL-JRNL-415990-DRAFT.

# I. INTRODUCTION

Gas proportional ionization counters (GPIC) have been used since the 1960's to obtain X-ray spectra [1]. These detectors rely on high electric fields that cause ionized electrons to induce a multiplying electron cascade, which is subsequently read out via charge-integrating electronics. To improve the energy resolution of the GPIC, gas proportional scintillation counters (GPSC) were developed, where the high electric field induces a greater number of scintillation photons than the GPIC induces cascade electrons, with an accompanying improvement in statistical accuracy. GPSCs themselves were improved by creating scintillation light in a smaller volume. This smaller volume, alternatively referred to as the amplification volume or gain region, allowed for a better-defined electric field, resulting in more uniform scintillation than would be available in a larger volume [2].

This proportional scintillation effect is obtained using both planar and cylindrical high-electric-field geometries [3]. In the planar geometry, a target volume is constructed with an electric field too low to create scintillation light, but through which ionized electrons drift toward an amplification volume. The amplification volume is close to a photomultiplier tube and has a uniform electric field high enough to create scintillation. In this way, possible systematic effects related to where the primary energy deposition occurred, such as light collection or electric field distortion, are minimized.

It is well known that a nuclear recoil induces less ionization in a noble gas or liquid than an electron recoil of the same energy. For many applications, particularly those that depend on particle identification, this effect must be accounted for to obtain correct results. The ratio in the observed energy between nuclear and electron recoils is referred to as the "nuclear quenching factor", and has been measured for various substances at different energies (see [4] and references therein). As the desired sensitivity of GPSCs increases, the requirement of measuring the nuclear quenching factor at lower energies becomes more important.

We have constructed a planar proportional scintillation detector using an argon/nitrogen mixture as the active medium to measure the nuclear quenching factor at an argon recoil energy of ∼8 keV. We used a $^{55}$Fe X-ray source to provide electron recoils, both as a means of energy calibration and to study the systematic behavior of the detector. Once the systematic effects are understood, the detector will be optimized for nuclear quenching factor measurements.

Some of the results of the current study apply to optimization of dual-phase detectors, in which the noble gas target region of the GPSC has been replaced with a liquid target volume. Because of their high target density compared to gas detectors and low energy thresholds, dual-phase detectors are attractive for use in the search for Dark Matter and coherent elastic neutrino scatter detection [5, 6, 7, 8]. Since the amplification volume of our detector is nearly identical to that of a dual-phase detector, much of our work is relevant for understanding the performance of dual-phase gain regions.

In the next section we describe the construction and operating conditions of our single-phase detector. In Section III, we describe the digital data acquisition system and our analysis of the raw digitized signals. In Section V, we examine factors that degrade the signal, and in Section IV we discuss the systematics of scintillation light production that are directly related to the physical setup of the detector. Section VI compares experimental and simulation data in the low-energy recoil regime. Throughout the article, we discuss ways to increase or stabilize the light generation or collection. We conclude with a discussion of ways to optimize detection of coherent neutrino scatter events.

## II. THE SINGLE-PHASE DETECTOR

In our single-phase detector, shown in Fig. 1, aluminum rings define the electric field in the drift region, and an aluminum hemisphere prevents breakdown between the high electrostatic potential elements and the steel chamber that enclosed the detector. The gain region was bounded above and below by fine wire meshes, which provide a more uniform electrical field than is present in the target volume. Each wire mesh was made of 30 $\mu$m gold-plated tungsten wires with a 1 mm pitch. The entire apparatus was supported on acrylic rods, with acrylic spacers providing physical and electrical separation where necessary. The field rings were 1 cm thick, separated by 1 cm, and their inner radius was 5.1 cm. The radius of the viewport was $\sim$3.8 cm, with a photomultiplier (PMT) radius of $\sim$2.5 cm. The gain region was 4 cm high, and the drift region was approximately 9 cm high.

The DC potential across the drift volume was established by linking adjacent field rings with 75 M$\Omega$ resistors. A 375 M$\Omega$ resistor was used between the upper, grounded grid and the lower grid nearest to the drift region. The rings farthest from the viewport were grounded to the aluminum hemisphere to create a region of zero electric field around the calibration

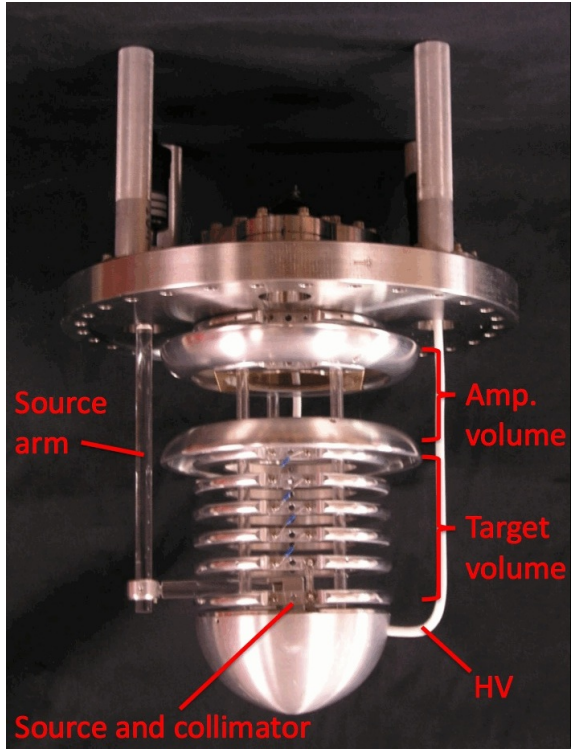

FIG. 1: Single-phase argon detector. The amplification volume is the open volume bracketed above and below by the high-potential field shaping rings. The target volume is below the amplification volume. The clear acrylic rod on the left is the pivot for changing the placement of the collimated source within the target volume. The curved white tube on the right is the HV supply for the drift volume inside a teflon insulator. The HV supply for the gain volume is partly visible through the gain gap, along with the three acrylic support rods.

source and collimator.

The rings farthest from the viewport, in direct electrical contact with the hemisphere, created a region of zero electric field around the calibration source and collimator. 75 MΩ resistors connected the remaining field rings, up to and including the drift-side grid. There was 375 MΩ of resistance between the drift-side grid and the grounded grid.

The detector itself was contained in a 40.6 cm × 40.6 cm stainless steel cylindrical pressure vessel. The temperature was not actively regulated. We evacuated the chamber and gas feed using a turbo pump until we achieved a vacuum of approximately $10^{-6}$ Torr. We then sealed off the pump and introduced ∼10 Torr of argon gas to the chamber to increase the pressure to a range compatible with the Baratron pressure gauge. We introduced 6 Torr of nitrogen gas to radiate light at a wavelength compatible with the PMT, as pure argon scintillates at 128 nm, a wavelength too short to be observed by the PMT. After filling with 6 Torr of $N_2$, we filled with an additional 390 Torr of Ar (optimization of the nitrogen content is discussed in Section IV A). We filled the chamber using this multi-step process so as to allow timely diffusion of the nitrogen into the chamber. The argon and nitrogen were obtained by from the Lawrence Livermore National Laboratory gas facilities, who reported the purity was 99.998% and 99.995%, respectively.

To optimize the electric potentials, we independently increased both the drift and gain potentials until we observed consistent electrostatic discharge. We then reduced the potentials to reduce the rate of discharge to no more than approximately once per hour. We developed an algorithm to exclude rare pathological events, including electrostatic discharges, from the final datasets. The algorithm is described in Section III E. The final potentials we used were 8000 V across the gain gap (E = 2 kV/cm) and 3000 V in the drift region (E = 0.333 kV/cm).

A 30-$\mu$Ci $^{55}$Fe calibration source was placed in a collimator inside the electrostatic hemisphere at the bottom of the field rings. The collimator directed the X-rays parallel to the axis of the target volume, toward the amplification volume. The moveable arm allowed the source to be placed at any distance between 0 and 4.5 cm from the central axis of the drift region. The collimator position was measured using the gradations on the rotary feedthrough. There was an aluminum plate that could be used to shield the target volume from the source.

Several techniques might be employed to improve the performance of the detector. The total pressure and gas concentrations can be optimized for maximum light output. Eq. 1 shows an empirical equation used in [9] for the production of light from ionized electrons traversing a scintillating gas:

$$N_{ph} = \int \alpha \left( E(x) - \beta p \right) dx \qquad (1)$$

where $N_{ph}$ is the number of scintillation photons generated, $\alpha$ and $\beta$ are constants of any given gas, $E$ is the electric field strength, and $p$ is the pressure of the gas. This equation shows that scintillation light output increases with electric field and decreases with pressure. The threshold for electrostatic discharge, however, increases with gas density, resulting in higher achievable operating electric fields. In the case of a single-phase detector, the total pressure can vary over a large range, while in a dual-phase detector the pressure is constrained by the cryogenic conditions. To narrow the range of independent parameters, a future optimization program can be guided by the pressure constraints required of a dual-phase detector.

There are a few reasons to eschew nitrogen to radiate longer-wavelength light when operating noble gas scintillation detectors, related to their use in Time Projection Chamber (TPC) mode. In TPC mode the original energy deposition creates a primary scintillation signal, followed some time later by secondary scintillation resulting from the electron transit through the gain region. The time delay between the signals provides a measure of the Z position of the original energy deposition. The relevant discussion in [9] concerning nitrogen as a long-wavelength radiator relies in part on the longer scintillation time constant of an argon/nitrogen mixture relative to pure argon (500 ns, versus <100 ns [19]). If the detector is used in TPC mode, the longer time constant of nitrogen scintillation ultimately increases the uncertainty of position along the direction of the electric field. Since the anticipated energy range of coherent neutrino scatters does not induce enough primary scintillation light to act as a trigger for the time projection chamber, the detectors used in their search are insensitive to Z position. Uncertainty in Z is therefore not a concern, and we still have the freedom to use nitrogen. This freedom notwithstanding, future efforts may compare the efficacy of using, for example, xenon or tetra-phenyl-butandiene (TPB) as a wavelength shifter.

Light collection efficiency can be increased by a number of means, including using Teflon or Tyvek reflectors, coating the photocathodes with materials which shift the light to a wavelength better matched to the PMT absorption spectrum, employing index-matching optical coupling agents at the PMT viewport, or transferring the PMTs to inside the chamber.

### III. DATA ACQUISITION AND ANALYSIS

We used a Hamamatsu 6522 photomultiplier tube biased at -2600 V. The PMT signal was digitized using a GaGe CompuScope 14200 digitizing PCI card running in a Dell Optiplex GX270 computer. The card obtained samples at 200 MHz, and we utilized a 100 MHz bandwidth and a 14-bit depth. The DAQ software was developed locally using LabView.

In anticipation of operations with a pulsed neutron beam, we triggered the DAQ with an external, uncorrelated TTL signal from a Stanford Research Systems DG535 pulse gen-erator running at 4900 Hz. We recorded a 200-$\mu$s trace for each trigger. This independent trigger eliminated any bias that might have arisen from a data-triggered readout scheme, as compared to a self-triggered readout. The speed of acquisition was ultimately limited by the speed of the hard drive, and reached a maximum of 8 MB/s, or approximately 50 waveforms per second. The final live time was ∼1%, with a data run lasting between 2 and 10 hours (see Section III C). A sample trace is shown in Fig. 2.

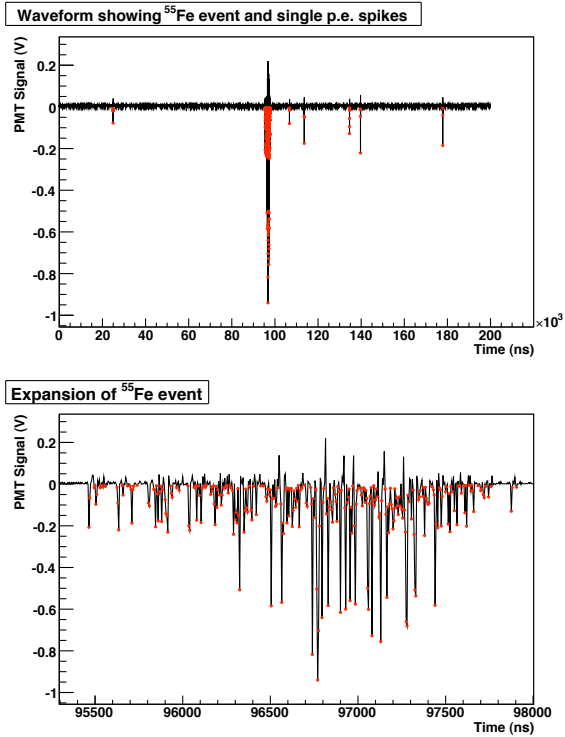

FIG. 2: A representative waveform acquired by the GaGe 14200 digitizer. A ∼2.5-$\mu$s-wide event, typical of an event created by electrons transiting the gain volume, is plotted in the top figure. The same event is plotted in the bottom figure on a finer time scale, showing the clear separation of single photo-electron pulses within the cluster. The red points show where the waveform exceeds the analysis threshold, with all other points being part of the baseline. The threshold was set at $3\sigma$ from the centroid of the baseline values.

### A.  Event characteristics

When an energy deposition occurs in the target volume, electrons are liberated from their atomic orbits. These ionized electrons drift through the low electric field region to the gain gap, where they subsequently induce scintillation in the argon gas. The wavelength of argon scintillation is 128 nm, well below the optimal wavelength sensitivity of the PMT. The nitrogen gas shifts the scintillation light to 337 nm [10], where its efficiency for detection is higher [11]. This shifted light is collected by the PMT, which puts out a short pulse of 10-20 ns for every detected photon. These pulses are emitted as the electrons transit the gain gap. This roughly two-microsecond period of scintillation is used to define an ionization event, consisting of the passage of one or more ionized electrons through the amplification region.

To assign the multiple PMT pulses to a single event, we defined a gate time, $T_g$, an interval between successive PMT pulses that exceed the analysis threshold. If an above-threshold pulse arrived within a time $T_g$ of the previous pulse, it was considered to be part of the same ionization event. The arrival of any pulse reset the analysis clock, so that the last pulse of an event did not have to arrive within $T_g$ of the first pulse in an event envelope, but rather within $T_g$ of the immediately preceeding pulse. Thus if $T_g$ was set too large for a given PMT dark rate, the total event width could become artificially long, up to the length of the entire 200-$\mu$s trace. If $T_g$ was set too low, a single event might have been divided into two or more sub-events, with a corresponding division of integrated energy. The end result is shown in Fig. 3, where combining a decreasing $T_g$ with a minimum event width of 2 $\mu$s increased the effective energy threshold. The curves in Fig. 3 were created by histogramming the integrated event envelope.

While increasing the gate time allowed for more complete signal collection and integration, it also increased the background rate. PMT pulses unassociated with an ionization event would fall within $T_g$ of the previous pulse, and extend the event envelope. This effect is shown in Fig. 4. With a longer $T_g$, it was possible to reduce the effects of unassociated pulses by requiring the event width be between 2.5 and 3.5 $\mu$s, but only at the cost of rejecting some valid events.

Thus we have demonstrated that increasing the gate time lowered the effective energy threshold, but at the cost of background rejection at the lowest energy. With a typical area of 1.5 V·ns, individual accidental pulses would enhance the 5.9 keV $^{55}$Fe X-ray peak by roughly 0.6%, and a 1-keV energy deposition would be enhanced by 3.5%. The total systematic uncertainty depends on the dark rate of any given PMT. With an envelope width of 2 $\mu$s and a dark rate of 20 kHz, the rate of a single accidental coincidence is $\sim 1\%$, so the final effect is very small compared to the resolution of the detector (see section III B for a measurement of the detector resolution). Since these systematic effects were small compared to the resolution of the detector, we decided to use a relatively long $T_g$ of 2 $\mu$s.

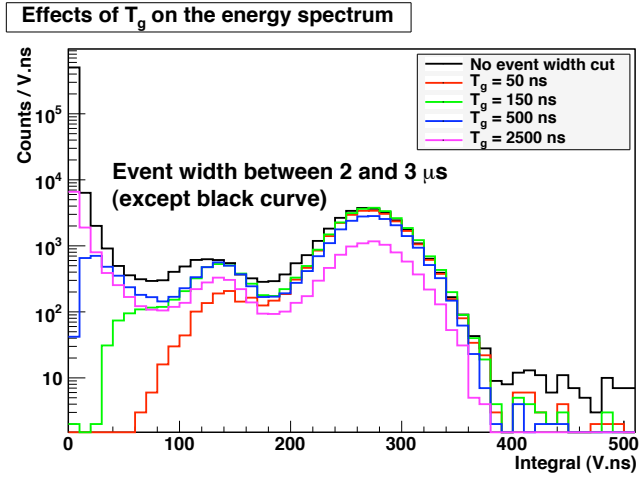

FIG. 3: Effects of gate time on the final spectrum. The longer the gate time, the lower the event energy that survives the event width cut. For these plots, the event width was required to be between 2 and 3 $\mu$s. The amplitude of the composite primary $^{55}$Fe peak (near 270 V·ns) in the $T_g$ = 2500 ns curve is below the others because there are fewer events that fall between 2 and 3 $\mu$s (see Fig. 4). Data was taken with the $^{55}$Fe calibration source in place.

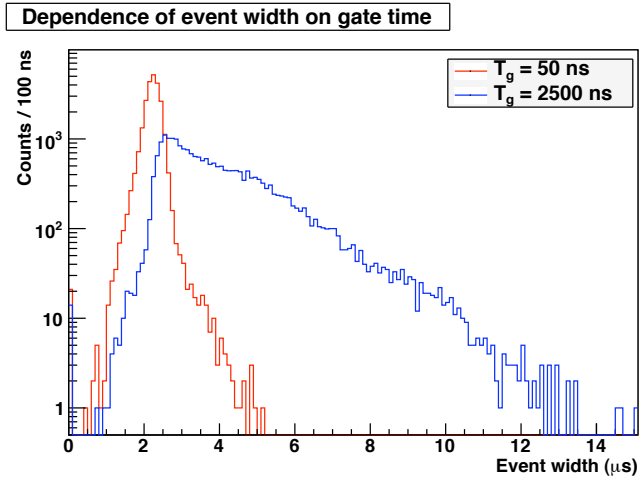

FIG. 4: The effect of gate time on the event width. The shorter the gate time, the fewer accidental single photoelectron pulses are included in the envelope, and thus the event width is both minimized and made sharper. These curves include an analysis cut around the composite primary $^{55}$Fe X-ray peak.

TABLE I: Most intense X-rays from $^{55}$Fe. The most intense X-ray not in this Table has a relative intensity of 0.28%, roughly 100 times smaller than the intensity of the combined 5.9 keV X-rays. This data was reproduced from the Table of Isotopes [12].

| Energy (keV) | Rel. Int (%) |
|:---:|:---:|
| 5.888 | 8.5 |
| 5.899 | 16.98 |
| 6.490 | 1.01 |
| 6.490 | 1.98 |

TABLE II: Electron transitions in argon given a missing K-shell electron. Transition energies and intensities come from the Evaluated Atomic Data Library [13].

| Transition | Energy (keV) | Frac. Prob. of Transition |
|:---:|:---:|:---:|
| K-L2 | 2.9282 | $3.559 \times 10^{-2}$ |
| K-L3 | 2.9305 | $7.033 \times 10^{-2}$ |
| K-M2 | 3.1630 | $2.913 \times 10^{-3}$ |
| K-M3 | 3.1632 | $5.756 \times 10^{-3}$ |

### B. Calibration and resolution

To calibrate the detector, we used both the primary X-rays from the $^{55}$Fe source as well as the X-ray escape peaks resulting from atomic transitions in the argon gas. The most intense X-ray lines from $^{55}$Fe decays are shown in Table I. For calibration purposes we assumed a weighted average of the most intense two, and combined the intensity of the last two to obtain 5.895 and 6.490 keV, with relative intensities 25.4 and 2.99, respectively. We performed this averaging to reduce the number of peaks in the calibration fit.

We also calculated the energy depositions following ejection of an argon K-shell electron by an incoming $^{55}$Fe X-ray. Table II shows the principal transition energies and intensities for the argon K shell. Repeating the above primary X-ray exercise for the transition probabilities within argon gives energies of 2.930 and 3.163 keV, with fractional probabilities 0.106 and 0.009, respectively.

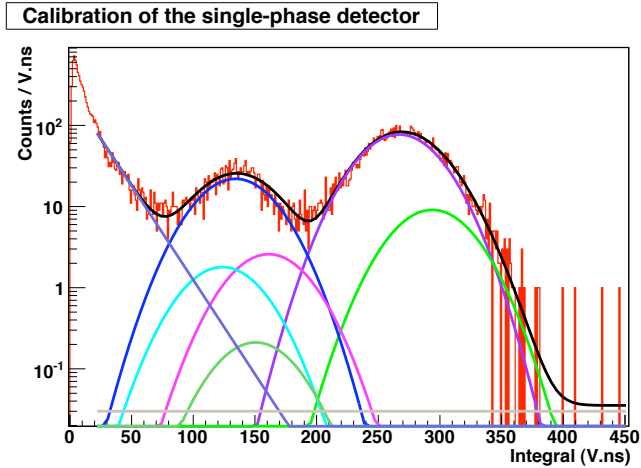

FIG. 5: Calibration of the single-phase detector. The data is shown in red, and the fit model in black. Each component of the fit model is shown by the rest of the curves. $\chi^2$ / d.o.f = 456.9 / 471, for a p-value of 0.67.

Each of the two primary X-ray peaks had two accompanying X-ray escape peaks for a total of six peaks, each of which is modeled by a Gaussian. We also assumed a constant plus exponential background, with the exponential curve having its own horizontal offset for ease of setting the initial values in the calibration fit. Thus there were originally 22 parameters in the calibration fit. After applying appropriate constraints on the peak intensities, energies, and widths, there were just nine free parameters left.

We applied the resulting calibration model to the prefiltered background data (see Section III C), with the result shown in Fig. 5. The reduced $\chi^2$ implies an acceptable calibration model. The calibration constant was 45.27 ± 0.07 V·ns/keV.

### C. Data prefilter

For the current study, we were interested in the events related to electron transition through the gain region. Most of the externally-triggered 200-$\mu$s traces, however, contained only single photoelectron (s.p.e.) pulses resulting from emission of thermal electrons within the PMT. We therefore prefiltered the data before recording it to disk, which conserved hard drive space and reduced analysis time.

In this prefilter, we required that at least one event in the trace have a width appreciably greater than that of an s.p.e. pulse. The width of a single pulse was up to 30 ns, and a small electronic reflection caused a secondary bipolar pulse about 40 ns later. While the amplitude of the reflected bipolar pulse was roughly 1/5 of the amplitude of the primary pulse, it diminished the total area by roughly 7%. For a relative measurement, however, the effects of this reflection were eliminated during detector calibration. To avoid the effects of the reflected pulse in the triggering, however, we set the prefilter timing threshold to 100 ns. If there was even one event in a waveform that had a width of at least 100 ns, the DAQ recorded the entire 200-$\mu$s trace to disk. Note that the DAQ trigger was unchanged between prefiltered and unfiltered data–the only difference was whether or not the trace was recorded.

As a systematic check on the prefiltering, we obtained both unfiltered and prefiltered data and plotted the resulting spectra on the same set of axes. The unfiltered and prefiltered data sets were accumulated over 2 hours (120 data files) and 9.5 hours (70 data files), respectively. There were 5470 counts in the composite primary X-ray peak in the unfiltered data and 28300 counts in the same composite peak in the prefiltered data, so the ratio of events was commensurate with the ratio of the length of the data runs. Given the unequal number of iron events in the two sets, we normalized each curve by the area of the primary 5.9 keV X-ray peak. Figure 6 shows the results after calibration and normalization. The two sets of centroids, amplitudes, and widths of primary and escape X-ray peaks are all within 1$\sigma$. We concluded the prefilter does not significantly affect the data given our set of analysis cuts.

### D. Relating event width and event integral

As an extra tool for analysis of the experimental data, we created a scatter plot to better understand the relationship between energy and event width, shown in Fig. 7. A cut around an energy of 6 keV preferentially selects true events resulting from electron transit through the gain gap. Conversely, an analysis cut around an event width of $\sim$2.3 $\mu$s will select $^{55}$Fe events, e.g., for calibration purposes.

The time length of the 3- and 6-keV peaks result from two effects. First, the continuous slowing down approximation range of a 3-keV electron in our chamber is $\sim$2.7 mm [14]. Using

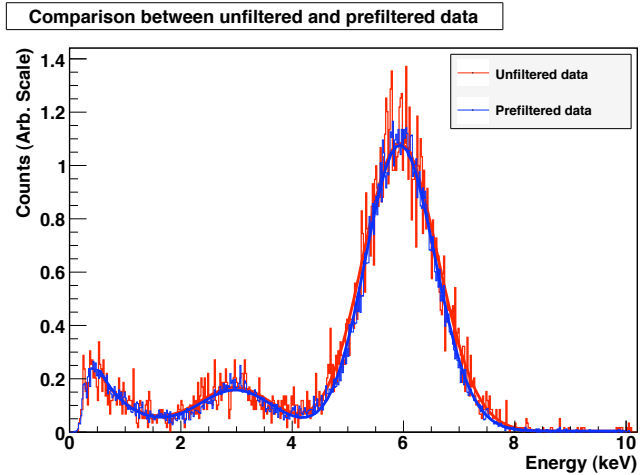

FIG. 6: Comparison of prefiltered and unfiltered data. The differences between spectral shapes are minimal. The prefiltered data set contained more iron X-ray events than the unfiltered data set, leading to smaller statistical fluctuations in the histogram. Graph available in color in online version.

*magboltz*, a Monte Carlo-based package to calculate electron transport speeds in gaseous environments [15], this 2.7-mm range translates to an exponential time constant of 0.12 $\mu$s. Second, the choice of gate time artificially extends the event envelope in the manner discussed in Section III A. Given the time characteristics of the tails in Fig. 7, the latter effect is dominant.

### E. Signal pathologies

The prefiltering DAQ occasionally recorded spurious events. A few examples of these traces are shown in Fig. 8. The highly differentiated signal shapes in these figures imply differing underlying causes. Possibilities include microdischarges, large discharges with a direct line of sight to the PMT, and inconsistent power supply to the analog electronics.

Instead of attempting to identify and characterize all the possible detector pathologies, we focused on identifying pathology-free traces. We hypothesized three possible parameters to categorize traces: the number of times the trace crosses the analysis threshold, the number of events in a trace, and the total time beyond analysis threshold (for a definition of the analysis threshold, see Fig. 2).

We started our pathologies analysis by integrating over the entire 200-$\mu$s traces from the

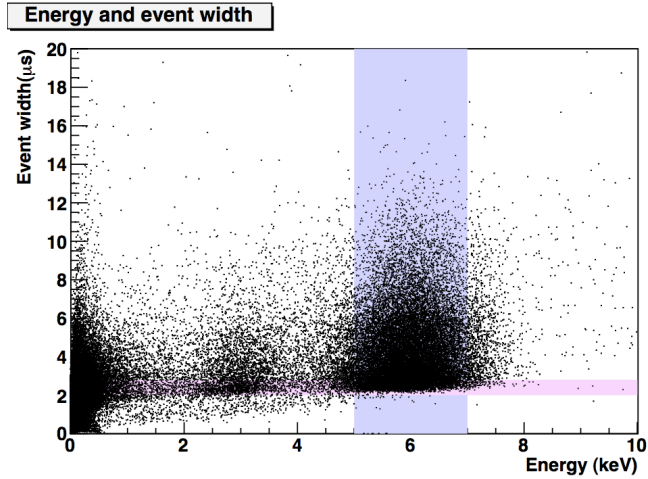

FIG. 7: Scatter plot of the event width and the event energy. The concentrations near 6 keV and 3 keV are the primary and escape X-ray peaks, respectively. The vertical band was used as an analysis cut for iron X-rays, and the horizontal band was used to select events caused by secondary scintillation in the gain region.

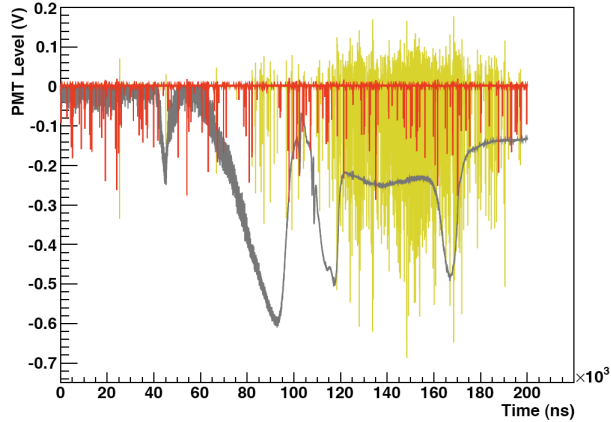

FIG. 8: Pathological detector traces. Each of the three traces (red, yellow, and grey) is an individual pathology. Compare to the normal-operations trace shown in Fig. 2.

9.5-hour "prefiltered" data set, and creating a histogram out of the integrals. The results are shown in Fig. 9. From this graph, we observed the expected primary and escape X-ray peaks near 140 and 280 V·ns, respectively. We also identified a higher-energy peak with a centroid near 560 V·ns, which comes from traces containing two iron events. The ratio of the area of this double-event peak to the area of the primary single-event peak is 0.6%, which compares favorably to the anticipated accidental coincidence rate of 1-2%. We therefore tailored our pathology analysis assuming no more than two $^{55}$Fe X-ray events in any given trace. This assumption will eliminate approximately one acceptable trace in 30,000.

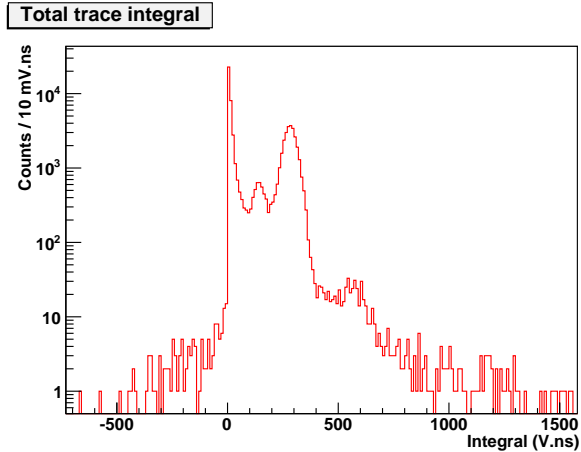

FIG. 9: Total trace integral of the 9.5-hour "prefiltered" dataset. The tail of the integral extends to over 8000 V·ns.

Figure 10 shows the number of times the trace crosses the software threshold, both for all traces and for traces with a single event from the $^{55}$Fe calibration source. The peak in the $^{55}$Fe curve has a centroid and deviation of 116.3 and 14.4 counts. Assuming a maximum of two iron X-rays in a trace, we adopted an upper limit on the number of crossings in an acceptable trace of twice the centroid plus three standard deviations, or 276. This analysis cut eliminated 0.5% of the traces. Note that we do not apply a lower limit on the number of threshold crossings, as that could eliminate lower-energy events that may be of interest in production analysis.

We histogrammed the number of events in a trace, and obtained Fig. 11. As with Fig. 10, we developed two curves, one for all traces and one for traces with a single iron X-ray. In this case, an "event" refers to any envelope of pulses as determined by the gate time analysis detailed in Section III A. The average value of the Poisson distribution fit to the $^{55}$Fe events in this figure is 4.65. To retain 99% of acceptable traces with a single iron X-ray, we would accept traces with a multiplicity of 11 or less. We allow for one additional iron X-ray in the trace, and thus we set our multiplicity cut to 12 or less. This cut eliminates 0.4% of traces from the data set.

Finally, we turned our attention to the total time the trace is beyond threshold. Figure 12 shows the total time a 200-$\mu$s trace extends beyond threshold. We use this curve to establish an upper limit on a single iron event of 8320 ns. We set this analysis cut to twice that value to allow for two independent $^{55}$Fe events, and therefore require a 200-$\mu$s trace spend no more than 16.64 $\mu$s beyond threshold. This 16.64-$\mu$s cut eliminates just 0.1% of the traces from the data set. As with the number-of-crossings cut, we do not apply a lower bound so as to avoid eliminating lower-energy events.

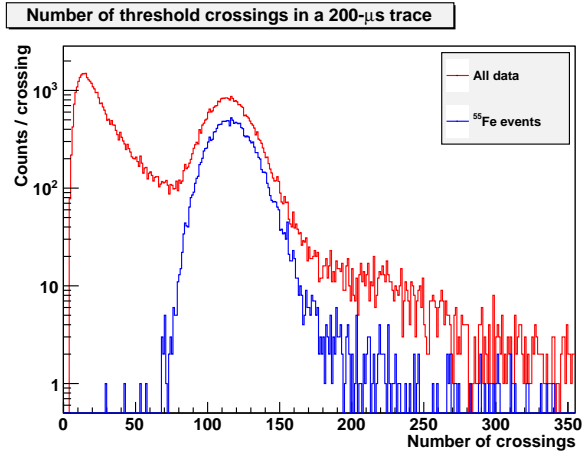

FIG. 10: Number of threshold crossings. The waveform crosses the threshold an average of ∼116 times in a regular $^{55}$Fe event. The number of crossings for some events extended to greater than 5000 in a single 200-$\mu$s trace.

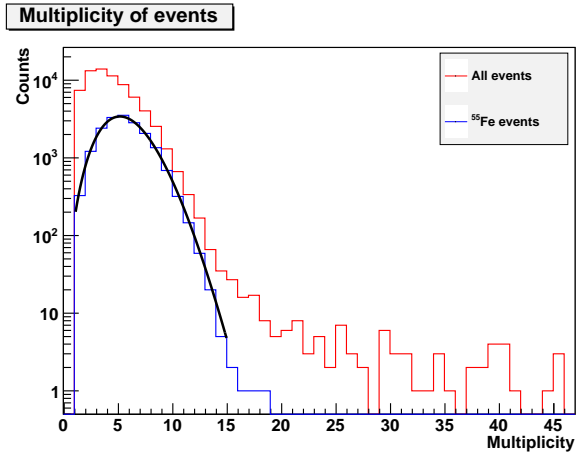

FIG. 11: Multiplicity of events in a 200-$\mu$s trace. The fit on the $^{55}$Fe curve is a Poisson distribution with $\lambda = 4.65$.

Applying the three pathological traces cuts to the data set eliminated 0.73% of the traces. Figure 13 shows the final spectra of events with all three analysis cuts, as well as the spectrum of events whose traces were eliminated by the cuts. Approximately 0.1% of the iron X-ray events have been eliminated by the cuts. This shows that our analysis cuts have been effective in eliminating spurious traces from the data set.

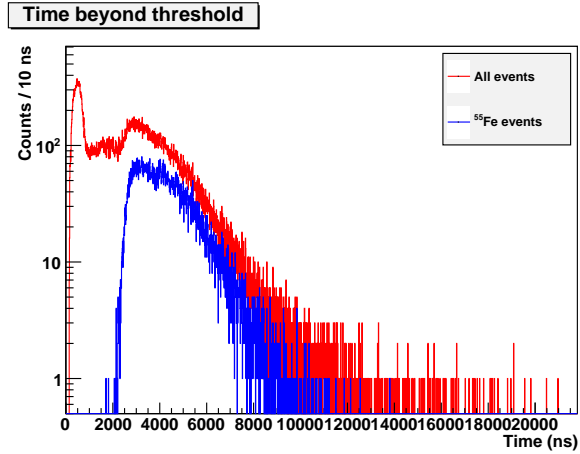

FIG. 12: Total time beyond threshold. Setting an upper limit of 8320 ns preserves 99% of single-X-ray traces. The tail of the red curve extends to 200 $\mu$s, the entire width of the trace.

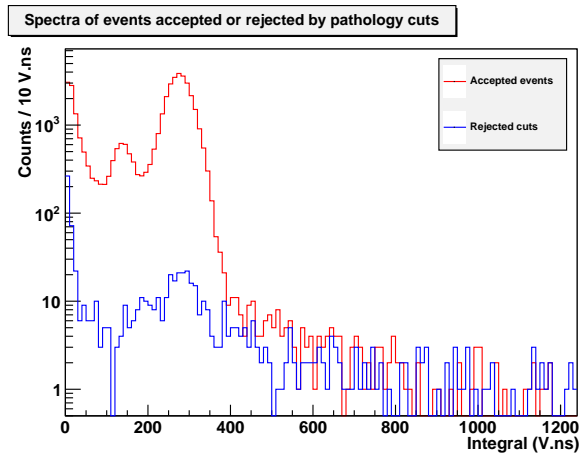

FIG. 13: Spectra of events accepted by pathological-trace cuts, and those rejected by the cuts.

## IV.  LIGHT PRODUCTION AND DETECTION

While it is possible in principle to observe both the primary and secondary scintillation within a single-phase detector, our detector's single viewport near the amplification region combined with the low energy depositions did not allow us to observe the primary scintil-lation. In this section we focus on the impact of the gas mixture on the secondary light production, as well as the optical response for off-axis energy depositions.

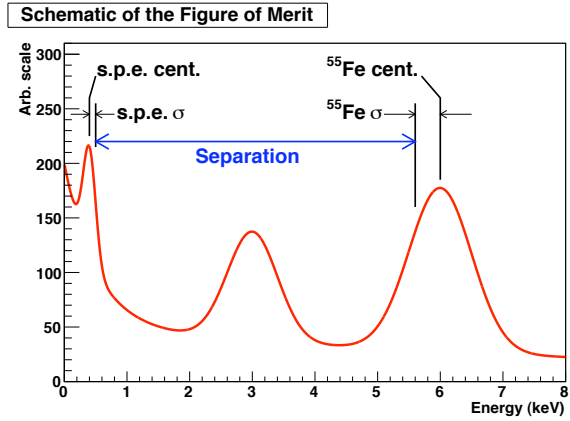

FIG. 14: Figure of Merit for characterization of quality of light collection. The F.O.M. itself is the separation, shown above, divided by the sum of the two sigmas, also shown above. The curve in this figure is not based on real data, but was constructed solely for illustrative purposes. In the actual experimental data, the s.p.e. peak is so narrow relative to the $^{55}$Fe peak that it would be vanishingly thin relative to the iron peak.

## A. Optimization of gas content

We adjusted the nitrogen doping of the detector to improve the light yield. To quantify the degree of improvement, we used the following figure of merit (F.O.M.):

$$F.O.M. = \frac{(C_{Fe} - \sigma_{Fe}) - (C_{spe} + \sigma_{spe})}{\sigma_{Fe} + \sigma_{spe}} \qquad (2)$$

$C_{Fe}$ and $C_{spe}$ are respectively the centroids of the primary $^{55}$Fe and s.p.e. peaks, and $\sigma_{Fe}$ and $\sigma_{spe}$ are their standard deviations. Figure 14 shows the components of the F.O.M.

Keeping the total pressure between 400 and 410 torr, we varied the nitrogen content from 0.75% to 2.75% in steps of approximately 0.25%. For each step, we recorded a spectrum and calculated the F.O.M. We found an optimal setting for a nitrogen gas content of ∼1.5% (see Fig. 15). Other experiments have studied nitrogen content in Ar/N$_2$ mixtures for 337 nm light production both below [10] and above [19] this value. The curve defined by the data in Fig 15 is somewhat broad, however, and while we are able to define an optimal nitrogen content, the relative standard deviation of the curve is ∼80%. Our results are therefore compatible with the cited references.

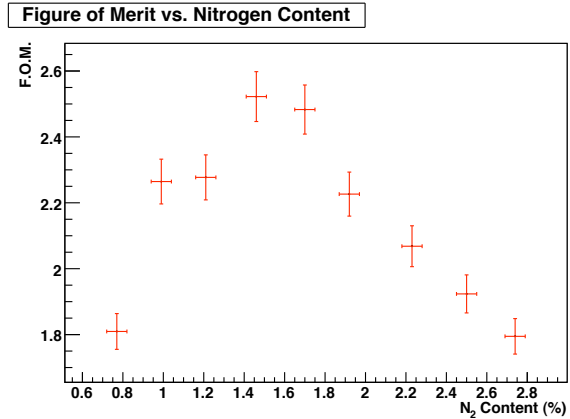

FIG. 15: Figure of Merit vs. nitrogen content. We found an optimal separation between the primary $^{55}$Fe 5.9 keV X-ray peak and the s.p.e. peak at an $N_2$ content of 1.5%. This is consistent with previous optimization studies of $N_2$ doping in Ar. See reference [10] for a discussion of the effects of nitrogen content on 337-nm scintillation.

To improve the nitrogen content optimization, it may be possible to perform an in-situ measurement of the gas content. One method would be to use a residual gas analyzer (RGA). This method would require additional plumbing, however, as a typical operating pressure for RGAs is on the order of $10^{-5}$ Torr. To use an RGA our system would have to be capable of withdrawing a representative sample of gas, rarifying it, exposing the RGA to it, and then evacuating the RGA chamber in preparation for further measurements. A second possibility that requires far less hardware would be to use the drift speed of the electrons within the gain region to determine gas content. Such an analysis is planned for a future study.

### B. Detector optical efficiency

We developed and validated a model of the optical response of the PMT as a function of radial distance away from the central axis of the detector. Based on the solid angle subtended over the PMT window, the probability of Fresnel reflection, and the attenuation of the light through the quartz glass, we created an analytic response curve as a function of radius, with physical parameters such as index of refraction and absorption coefficients fixed.

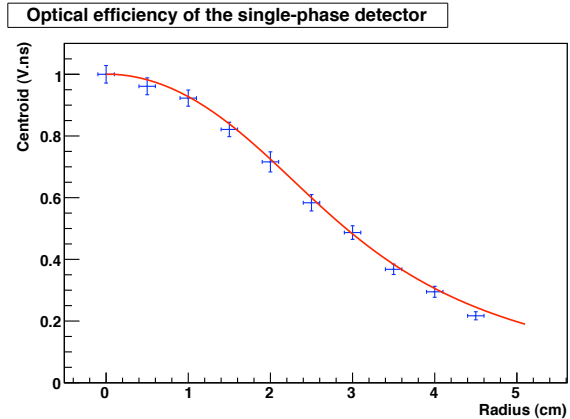

FIG. 16: Optical efficiency of the single-phase detector. The experimental data points were obtained using a moveable, collimated $^{55}$Fe source. Note that the solid curve is an independent, analytical prediction, rather than a best fit to the data.

To verify our optical model, we obtained data with the collimated $^{55}$Fe source placed at radii from 0 to 4.5 cm in 0.5 cm steps. We plotted the data along with our optical model in Fig. 16. The data supports our optical model, which we applied to our simulations (see Section VI).

## V. TIME-VARYING RESPONSE

Over the course of our investigations we discovered that the detector response to the continual $^{55}$Fe source was variable over time. The 5.9 keV peak centroid diminished and approached an asymptote. We observed discontinuities in the asymptotic decline both between data sets as well as during a run itself. Figure 17 shows the signal degradation and fluctuations. In this Figure, we plot the uncalibrated centroid of the iron peak for the background and subsequent gamma data sets discussed in Section VI. For the gamma data sets, we kept the $^{55}$Fe source in place to provide a constant energy calibration, and we simply placed various gamma sources on the side of the detector chamber. We observed discontinuities in the background data as well as between the $^{137}$Cs and $^{60}$Co data sets.

Immediately after the background data run, we evacuated the chamber for approximately two days, and refilled with fresh argon and nitrogen. We allowed an extra day and a half to pass to allow the system to settle before acquiring the gamma data. We hypothesize that the gradual degradation in the centroid during the $^{137}$Cs data run arose from water and $O_2$ outgassing from the detector materials. These electronegative molecules absorb ionized electrons as they drift through the target volume, resulting in less light production.

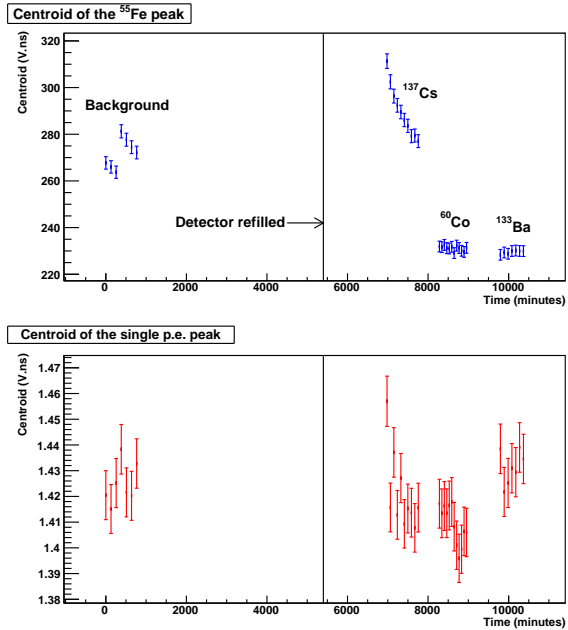

FIG. 17: Stability of the detector response. We observed two discontinuities in the upper graph, within the background data set and between the $^{137}$Cs and $^{60}$Co data sets. Sparking may have caused these discontinuities, though we have not conclusively proven this hypothesis. The lower graph shows the centroids of the s.p.e. peaks, showing consistent results from the PMTs, electronics, DAQ, and analysis. For both graphs, the error bars show the width of the peaks, not the uncertainty of the peak centroids. The horizontal error bars are too small to be seen on this scale.

Although there is not yet conclusive proof available, the discrete increase in detector response shown in Fig. 17 during the background data may be due to shifting nitrogen content in the gas resulting from electrostatic discharge. In [16], Matsuda *et al.* relate time-varying light output to nitrogen plating on electrodes in a gas discharge lamp. In our case, after quickly plating on the electrodes the nitrogen would slowly desorb from the electrodes, returning the detector response to the same level as before the electrostatic discharge.

As a systematic check on the detector response, we measured the centroid of the single photoelectron peak, and with the exception of an outlier in the $^{137}$Cs data set, the variations in the centroids are roughly on scale with the uncertainties (Fig. 17). We therefore conclude that the changes in the response to the $^{55}$Fe source were related to changes in either the gaseous medium or high-voltage supplies, and were not based in the PMTs, digitizer, LabView DAQ, or analysis routines. The grid high-voltage supplies, while occasionally undergoing electrostatic discharge, not only recovered in less than a second, but showed no long-term variations. We conclude the time-varying detector response is primarily related solely to the gaseous medium. To correct for the drift in the detector response, we re-analyzed all data with a re-calibration performed approximately every hour. The data presented in this work includes this hourly calibration.

The variable detector response detailed in this section might be made more constant by a gas purification system. This purification system must not only eliminate the major atmospheric contaminants ($O_2$, $H_2O$, $CO_2$, etc.), but it must not eliminate $N_2$ [17], nor should it introduce radioactive backgrounds [18]. Another factor that may be contributing to the signal variation is temperature fluctuations. These effects are now being addressed in an upgrade to the detector described in this work.

## VI. COMPARISON WITH SIMULATION

The first scientific objective for this detector is to measure the nuclear recoil quench factor in the energy range of 1-10 keV with a neutron beam of appropriate energy. This beam may have source-associated gammas, both prompt and delayed, which cannot be removed with fiducial cuts. We would need to simulate the effects of both these source-associated gammas as well as background gammas so as to account for them in the final dataset. To this end, we tested the validity of a few simulation codes and models.

We created a GEANT4 simulation [20] of the detector that incorporated the stainless steel tank wall and flanges, the aluminum ring and field shapers, and the aluminum hemisphere enclosing the collimated source (see Fig. 18). The simulation made use of the G4LowEnergy electromagnetic models for gamma, electron, and positron interactions, and we used the GEANT4 Radioactive Decay Manager to create decay particles from a collimated $^{55}$Fe source.

We convolved the simulated energy depositions with both the optical response and the experimentally-measured finite detector resolution. We also attenuated all energy depositions that occurred within the gain region by an amount proportional to the distance from the location of the deposition to the top of the gain region, e.g., if the energy deposition occurred 1 cm from the top of the 4 cm gain region, the energy deposition was reduced by 75% (as a point of reference, the attenuation length of 5.9 keV X-rays in our gaseous mixture was approximately 2.2 cm). We normalized the simulation and experimental spectra by the areas of their respective 5.9-keV X-ray peaks, and plotted them in Fig. 19.

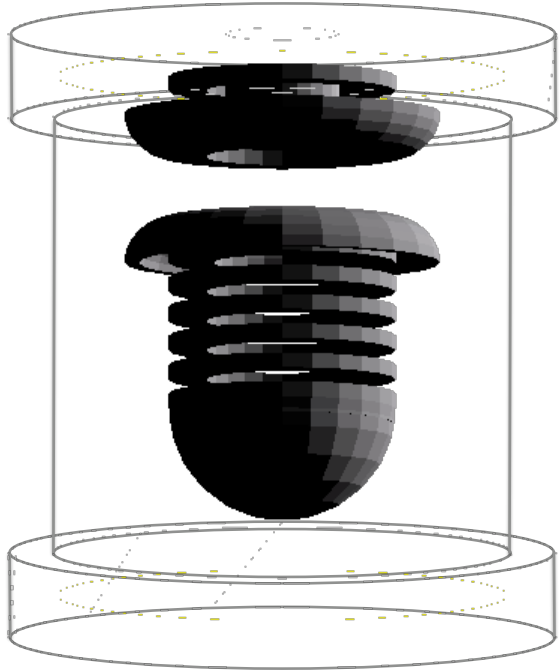

FIG. 18: GEANT4 simulation of the detector. the aluminum field-shaping rings and the hemisphere enclosing the collimated source. The pressure vessel tank is shown in wireframe for clarity. The experimental data which were used for comparison with the simulation were taken with a different detector geometry than that described in Fig. 1, and matches the geometry shown here.

There were clearly background events in the experimental data that were absent from the simulation. We found it necessary to run with the $^{55}$Fe source constantly present, as the calibration constant was fluctuating over time. As a result, we could not subtract the non-$^{55}$Fe contributions to the spectrum. Our primary point of validation, therefore, was the ratios between the primary and escape X-rays peaks. For the experimental data, the ratio of areas was $8.34 \pm 4.12$. For the simulation data the ratio was $10.69 \pm 3.02$. The uncertainties in these ratios were derived solely from the statistics of the fit routine, and were driven primarily by the uncertainty in the area of the escape peak and the underlying background model. The ratios were consistent to within the experimental uncertainties.

To test the simulation in a different photon energy range, we obtained three more ex-perimental data sets, using $^{60}$Co, $^{137}$Cs, and $^{133}$Ba. We placed the sources on the outside wall of the steel tank, roughly at the same height as the middle of the target volume. We acquired data from each gamma source for between 9 and 13 hours, and we calibrated the data approximately every 60-90 minutes. Simulations were then run for each of the sources, again using the GEANT4 Radioactive Decay Manager to create the primary particles. The simulated energy depositions were convolved with the optical model and attenuated as described earlier in this section. For the $^{60}$Co simulations, we normalized the simulation and experimental results to have the same level in the 16-20 keV energy range. Figure 20 shows the comparison of the simulation and experimental $^{60}$Co data.

We did not find agreement between the GEANT4 simulation and experimental results. We ran an MCNP [21] simulation as a systematic check on GEANT4. Additionally, we decided to explore alternative low-energy electromagnetic models within GEANT4. We reran the simulations using the G4LECS extension [22], and the G4Penelope [23] models. There was no appreciable difference between the various GEANT4 models (see Fig. 21). A similar discrepancy between simulation and experiment was observed when comparing the $^{137}$Cs and $^{133}$Ba data sets, although they were not as drastic as the differences in the $^{60}$Co data.

One possible source of the discrepancy is that the individual energy depositions are below the thresholds of both MCNP and GEANT4. For example, the GEANT4 collaboration claims accuracy down to 250 eV. While this limit is below where experiment and simulation diverge, it could be that events that deposit up to 3 keV total are made up of smaller energy depositions that are below GEANT4's capabilities.

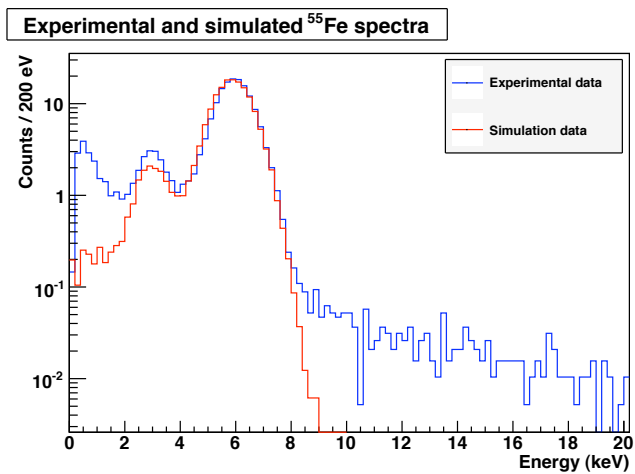

FIG. 19: Comparison between simulation and experimental data from the $^{55}$Fe source. The validity of the simulation was determined from the relative areas of the primary and escape X-ray peaks.

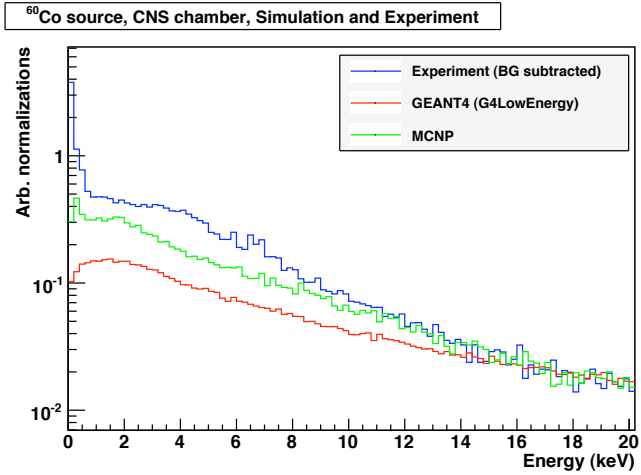

FIG. 20: Comparison of experimental and simulation $^{60}$Co data. The GEANT4 results did not agree with experiment, so we performed a second simulation using MCNP. The experimental and simulation results still did not match well, especially in the crucial low-energy region.

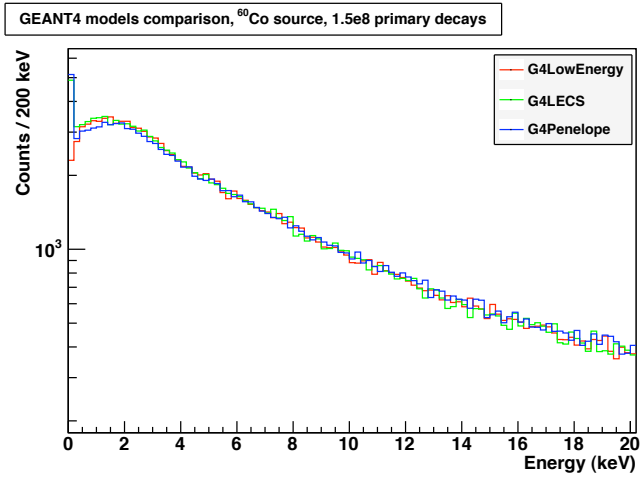

FIG. 21: Comparison of three different GEANT4 models. The same number of primary decays was used for all three simulations, and no further normalization has been applied to the results. The models did not show any appreciable difference, neither in the overall shape, nor in amplitude.

A second explanation could be the production of delta electrons in the experimental setup. While we calibrated our detector results to correspond to an energy spectrum, at a more basic level we were simply counting the number of electrons that enter the gain region. For the sake of illustration, assume a 6 keV energy deposition liberated 200 electrons. But if 200 delta electrons were created instead, each with an energy of just 1 eV, the event would be interpreted as a 6 keV event even though the original energy only summed to 200 eV. While qualitative, this illustration demonstrates how our calibration may break down at the quantum level. A fiducial cut may be able to reduce the effect of any delta electrons that originate from the aluminum rings, but with just a single photomultiplier we did not have a reliable method to determine the radial position of an energy deposition.

Whatever the source of the discrepancy, we were not able to validate the simulation for high-energy gammas. When analyzing data in the future, we must determine an alternate method to eliminate gamma backgrounds from nuclear recoils. Some background events from high-energy gamma rays can be identified through the use of fiducialization. We are interested primarily in the energy range up to 10 keVee. At that energy, an electron will travel roughly 7 mm through 400 torr of argon [24], and an argon nucleus will travel much less. If we can apply fiducialization to the detector, we will not only greatly diminish the effects of delta electrons, but we can also remove most high-energy Compton scatters, which usually result in electron energies much greater than 10 keV.

## VII. DISCUSSION & CONCLUSION

While dual-phase detectors are used or proposed for Dark Matter searches and the detection of coherent neutrino scatters, our primary focus is on measuring the nuclear quenching factor in gaseous argon. The energy range of interest is below 10 keVee. Since the energy deposition is so low, we are not concerned with identification of the primary scintillation light. This relaxes the proscription against using nitrogen as a long-wavelength radiator. Our gas pressure and content optimization procedures will take this into account, using as a starting point the measured optimum of 1.5% $N_2$ at 400 torr. We will adjust the grid potentials to be as high as possible without undue levels of electrostatic discharge.

We must maximize light output from these weak recoil-induced ionization events. To ensure maximum integration of the light signal, we will set the gate time to be the transit time of the electrons across the gain gap. While a final analysis cut on the event width must wait for nuclear recoil data, we anticipate using the gate time as the minimum required event width. We might also use an upper limit on the gate time that is less than the transit time across the full target volume. In Section III A we concluded that the effect of accidental single photoelectron pulses will not appreciably alter the final energy spectrum.

We will employ a gas recirculation and purification system to reduce the uncertainty in the final spectra. If the detector response can be stabilized, we can shutter the $^{55}$Fe source when acquiring nuclear recoil spectra to reduce the background levels. We will also continue to acquire data using a prefilter to increase the effective data rate. We will also use multiple, internal PMTs to allow for fiducialization and to increase light collection efficiency.

Dual-phase detectors have some great advantages when searching for rare, low-energy nuclear recoils. Understanding the gaseous phase of these detectors is crucial to experimental success. With the systematics outlined in this work and the improvements listed in the previous section, we hope to provide some guidance for measurements of basic physics.

## VIII. ACKNOWLEDGMENTS

We would like to thank Dennis Carr and Darrell Carter for their engineering and fabrication support. Thanks also to Sean Paling and Nigel Smith of the ZEPLIN-II collaboration for an explanation of the alpha backgrounds related to their SAES gas purifier. Thank you to Steven Dazeley for a thorough reading before submission for publication.